\documentclass[aps,tightenlines,prc,preprint,showpacs,showkeys,floatfix]{revtex4}
\usepackage{epsfig}
\newcommand{\bfDelta}{\mbox{\boldmath $\Delta$}}
\newcommand{\boldsigma}{\mbox{\boldmath $\sigma$}}

\newcommand{\bfP}{{\bf P}}

\def\etal{{\it et al.\/}}
\def\bfq {{\bf q}}
\def\bfs {{\bf s}}

\def\adagg{{a^\dagger}}
\def\bfl{{\bf l}}
\newcommand{\bdagg}{{b^\dagger}}
\def\bfK{{\bf K}}

\def\bfL{{\bf L}}
\def\bfR{{\bf R}}
\def\bfk{{\bf k}}
\def\bfp{{\bf p}}  
 
\def\bfr{{\bf r}}

\def\be{\begin{equation}}
 \def \ee{\end{equation}}
\def\bea{\begin{eqnarray}}
  \def\eea{\end{eqnarray}}
\newcommand{\eqn} {Eq.~(\ref )}
\newcommand{\bb}{\langle}
\newcommand{\kk}{\rangle}
\def\Z{{\Theta}}

\def\bfQ {{\bf Q}}

\def\bfq {{\bf q}}
\def\bfK{{\bf K}}
\def\bfL{{\bf L}}
\def\bfk{{\bf k}}
\def\bfp{{\bf p}}  
\def\be{\begin{equation}}
 \def \ee{\end{equation}}
\def\bea{\begin{eqnarray}}
  \def\eea{\end{eqnarray}}
\def\eqn {Eq.~(\ref }
\def\boldv{{\bf v}}

\date{\today}
\begin{document}
\vskip1in\vskip1in\vskip1in\vskip1in
\preprint{NT@UW-04-03,ECT-03-035,LBNL-54638}
\title{Even Parity $\Theta$-Pentaquark
and Stable Strange Nuclear Matter}

\author{Gerald A. Miller}
\affiliation{Department of Physics, University of Washington, Seattle, 
WA 98195-1560}

\begin{abstract}
The newly discovered exotic $\Theta$
 baryon of mass 1540 MeV (and very small width)
truly has a very low mass,  if it is a pentaquark system of even parity.
A schematic model in which the coherent interaction of 
$u\bar s$ and $d\bar s$ pairs leads to a very 
large residual (non-confining) attractive 
interaction is introduced. This collective vibrational model  accounts for the 
mass and small decay width to the $KN$ channel, but yields a significant
 coupling
to the virtual $K^*N$ channel.   
The schematic model predicts 
 an
attractive $\Theta$-nucleon interaction   strong enough 
to bind a 
$\Theta$  particle to a nucleus   in a  state that is stable against 
 decay via strong  interactions.
The discovery of $\Theta$-nuclei could be
a definitive proof that the $\Theta$  parity is even.
\end{abstract}


\maketitle

Stimulated by the prediction\cite{Diakonov:1997mm} of the existence 
of an extremely narrow positive strangeness pentaquark state $\Z$ of mass
 1.54 GeV and $J^P=1/2^+$, 
a member of an antidecuplet, 
several experimental searches were
 successfully undertaken\cite{Nakano:2003qx}-
\cite{Aleev:2004sa}.
Other possible  strangeness -2 partners in the $\Z$-antidecuplet 
were also detected\cite{Alt:2003vb}.

 A study of the implications of
the existence of the state with the quantum numbers 
predicted in Ref.~\cite{Diakonov:1997mm} is presented here.
An even-parity, low-lying  S=+1 state
must have very unusual properties\cite{Zhu:2003ba}, with 
 the most unusual
being is its very low mass. Indeed, 
lattice calculations\cite{Csikor:2003ng} find the expected result that 
the odd parity $S=+1$ state is the one of lowest energy.
Furthermore, the lack of a presence in
$KN$ scattering  seems to limit the width to an usual  small value (perhaps on
the order of an MeV)\cite{Cahn:2003wq}, 
but its
relatively large production cross section could arise from the exchange
of $K^*$ mesons\cite{Karliner:2004qw}. 
A state with such unique properties does not arise in an obvious way
from  previously  existing
quark models of baryon spectroscopy, so it would be 
interesting to create new models with new implications.

The unusual nature of the $\Z$ can be 
seen immediately using the naive quark model. 
The $\Z$ and nucleon have the same $J^P$,   their
mass  difference
  arises  from the addition of a pair of  constituent
quarks $d\bar s$ or $u\bar s$ (of mass 340 and 510 MeV)  and  
 one unit of orbital angular momentum  to the
nucleon.
is present.
The  energy cost of the unit of angular 
momentum can be estimated as  the 
mass difference between the nucleon and lowest-lying odd parity excited
states, or  about 600 MeV.
 Thus the naive quark model gives the energy of the $\Z-N$ mass
difference, $\hbar\omega_0$,
as 
about 1.4 GeV! 
This truly astonishing number makes it clear
that the naive quark model
does not contain the  interactions
 between quarks needed to explain low mass of  
 the $\Z$.
In the original  work\cite {Diakonov:1997mm}, the attraction was assumed to
arise from the non-perturbative effects of the chiral soliton
field.  Such a field can only 
have its quark model interpretation in terms of non-confining residual
interactions.
Indeed the aim of understanding the $\Z$ dynamics has attracted several 
proposed mechanisms, including:
 flavor and color hyperfine interactions (see the review in  
Ref.~\cite{Jennings:2003wz}), and 
strong quark-quark interactions leading  to di-quarks\cite{Jaffe:2003sg}
 (perhaps caused by
the influence of instantons in the vacuum\cite{Shuryak:2003zi}).
 We'll  provide another mechanism, perhaps similar to
 some of the others, to obtain  a wave function that
differs significantly from that of the naive quark model.

Here is an outline of our logic. We'll work in the framework of the 
quark model, so that some 
 gargantuan residual  interaction is needed. Since
 no ultra-large  coupling constants can be  expected to arise from QCD, the
   extreme strength must arise from  collective effects.
We introduce a schematic interaction that causes   the 
$\Z$ to be  described as a coherent set of color-singlet 
$d\bar s, u\bar s$ pseudoscalar excitations  that move in 
a p-wave relative to the  nucleon.
This   interaction
between these  excitations and the nucleon must be assumed to
be  attractive, but need not be very strong. In this case, the $\Z$ can be
regarded as a collective vibration of the nucleon. 
This mechanism is reminiscent of one used
to describe the nuclear giant dipole resonance\cite{bb}.
The model can account for the mass, width and $K*$ interactions of the 
$\Z$.  But the 
 most dramatic implication of the model is that 
$\Theta$ to binds to nuclei with a binding energy
 considerably greater than 100 MeV. This new   $\Z$-matter 
would therefore be  stable against
decay by the strong interaction. 

  It is worthwhile to stress that the
unusual implications we  draw stem entirely from the presumed 
even parity of the $\Z$. An unambiguous experimental determination 
of  the 
 $\Z$ parity would be very important, and   necessary 
experiments are being planned\cite{Thomas:2003ak},
\cite{Hanhart:2003xp}.

Let's turn to the calculation of the 
$\Theta$  wave function.  The goal is to model a effective interaction
between the pseudoscalar excitations and the nucleon
that reproduces the energy and decay properties of the $\Z$.
Consider  a  color-singlet 
$d\bar s$ configuration of even parity that moves in a relative
p-wave around the nucleon. 
The $d$ quark can have any one of three colors, and the pair 
can have any of three values of $L_z$, so there are 9
possible states.
 Similarly there can be 
$u\bar s$ configurations, so there are 
18 possible states. 
The  coherence  of 18 states
will  lead to considerable collective effects.
Furthermore, we can include radial excitations. In this case,
number of states will be unlimited. 

We'll
use a specific schematic model
to see how coherent effects can cause strong attraction. 
Let  $a_n(\bfk)$ 
denote the operator that destroys a pseudoscalar excitation in a discrete
level $n$, 
and let $b_{m}(\bfp)$ denote the nucleon destruction operator.
 Our proposed wave function for the  
$\Z$ state of total momentum $\bfP$ and spin $m$ is given by 
\bea
\vert\Theta(\bfP,m)\rangle=
\sum_n\int d^3q\;\adagg_n(\bfq)\sum_{{m}'}\langle m\vert
\boldsigma\cdot\bfq_r\vert m'\rangle \bdagg_{{m}'}
(\bfP-\bfq)c_n(\bfq_r)\vert0\rangle
,\label{wf}\eea
where $\bfq_r\equiv(M\bfq-\mu(\bfP-\bfq))/(M+\mu)$ 
is the relative momentum between the meson (of mass $\mu$) and nucleon
constituents of the $\Z$. The function $c_n$ will be determined
by solving the Schroedinger equation.
The wave function of \eqn{wf}) 
is a coherent superposition of states, and this coherence will  generate 
a huge
attraction. The normalization condition is 
$\langle\Theta(\bfP',m')
\vert\Theta(\bfP,m)\rangle=\delta_{m,m'}\delta(\bfP-\bfP').
$
We'll show below that using color-singlet pseudoscalar excitations in a 
 p-wave  relative to the nucleon
substantially suppresses   the decay to $KN$, while
allowing
the virtual transition to $K^*N$.

Now let's write the Hamiltonian.
  The  nucleon and $\Z$ are treated as heavy objects, and
the energy  carried by the the $i$'th excitation is denoted as
$\omega_i$, with 
\bea\omega_i=\mu +\Delta E_1+\Delta E_i, \label{oi}\eea
where the energy 
required by having a  relative p-wave is $\Delta E_1$, the energy
due to a possible radial excitation is $\Delta E_i$, and the energy
required by having 
two extra
quarks of total effect mass is $\mu$.
As noted above 
the naive quark  model gives   $\omega_0\approx1400 $ MeV, 
if radial excitations
are ignored.
The value of $\mu$ in the naive quark model would be about 
850 MeV, but this ignores the potential 
influence of an attractive confining and hyperfine interactions between
the $\bar {s}$ and its partner. 
This attraction could be about 600 MeV \cite{Isgur:xj}
so that  $\omega_i$ is at least aproximately 
900  MeV, and $\mu$ is at least 250 MeV.
Such energies are much larger than
the expected  kinetic  energies of the  pseudoscalar excitations and the
nucleon, so the latter are neglected.

Given the above definitions and aproximations, 
the unperturbed Hamiltonian $H_0$ can be  expressed as
\bea
H_0=M\sum_{{m}_s}
\int d^3p \;\bdagg_{{m}_s}(\bfp) b_{m}(\bfp) +\sum_i\omega_i\int d^3k\;
\adagg_i(\bfk)a_i(\bfk),\eea
where $M$ is the nucleon mass.
The residual interaction $\widehat{V}$ is chosen as 
\bea
\widehat{V}=
-\lambda\sum_{i,j,m_s}\int d^3kd^3k' d^3p\;\adagg_i(\bfk')\bdagg_{{m}_s}
(\bfp+\bfk-\bfk')
a_j(\bfk)b_{{m}_s}(\bfp)
D_i(k'_r)D_j(k_r)
\bfk'_r\cdot\bfk_r
,\eea
where $\bfk_r=(M \bfk-\mu\bfp)/(M+\mu),
\bfk'_r=(M \bfk'-\mu(\bfp+\bfk-\bfk'))/(M+\mu),$ and $
k_r=\mid \bfk_r\mid$
If the $\Z$ is in its rest frame ($\bfk+\bfp=0)$,  the relative momenta can be
taken to be that of the pseudoscalar excitation (or the negative of the
nucleon momentum). In this case, $\bfk_r=\bfk,\bfk'_r=\bfk'$.
 The real functions
$D_i$ are  to be defined as part of the model,
 with a spatial extent  determined by   
the nucleon radius, $R_0.$ 
The interaction $\widehat{V}$ involves all 5 quarks, and can arise through
two successive three-quark 't~Hooft
interactions proceeding via  the instantons in the vacuum. Thus three separate
flavors are required for this mechanism to occur.

\begin{figure}
\begin{center}
        \leavevmode
        \epsfxsize=0.7\textwidth
        \epsfbox{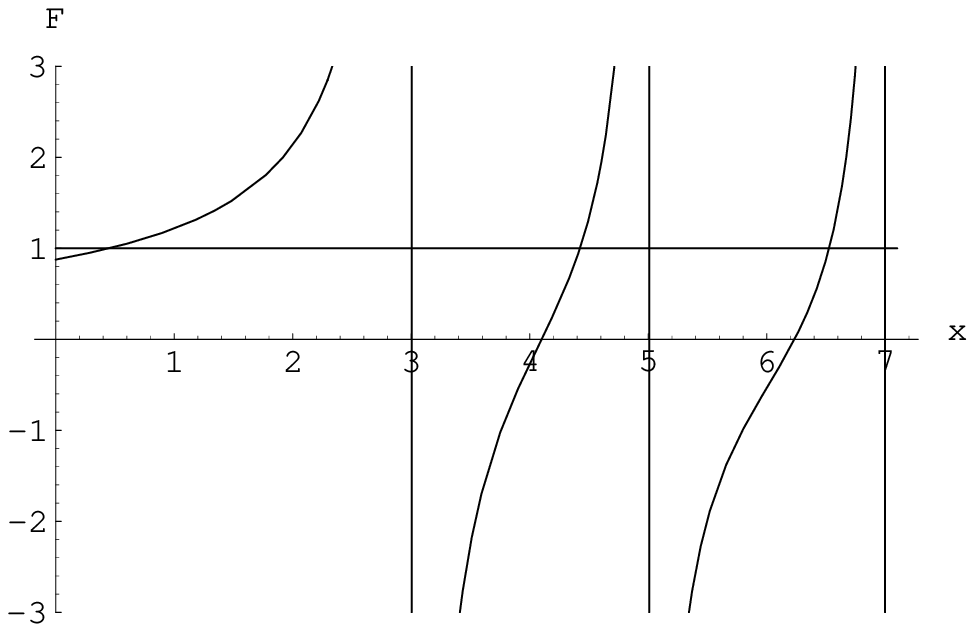}
\end{center}
\caption{\label{fig:trans1} Graphical solution of \eqn{trans0}).
The variable $x$ represents the quantity $M_\Z-M$.}
 \end{figure}

Proceed by using
 \eqn{wf}) (for $\bfP=0$) in the Schroedinger  equation with  the 
Hamiltonian $H=H_0+\widehat{V}$, and  then  acting
 with $a_j(\bfl) $, where $\bfl$ is an arbitrary momentum vector, 
on both sides. After some algebra, one obtains 
the result:
\bea c_j(l)=-{\lambda\over3}{ D_j(l)\over(M_\Theta-M-\omega_j)}
\sum_n
\int d^3q\; D_n(q)c_n(q)q^2.
\label{ceq}\eea
This relation is a consistency condition between the Hamiltonian and
wave function, and can be re-written as a transcendental equation
for $M_\Z$ using techniques developed long ago\cite{bb}. To see this, 
multiply \eqn{ceq}) by $l^2D_l(l)$,   
integrate over $d^3l$, sum over $j$ and divide by a common factor
to obtain:
\bea
1={\lambda\over3}\sum_j{ \int d^3l\;l^2 D_j(l)c_j(l)\over (-M_\Theta+M+\omega_j)
}.\label{trans0}\eea
 Consider the right-hand-side to be a function of $M_\Z,\; F(M_\Z)$ that
 has poles at $M_Z=M+\omega_i$, with $F(M_\Z)$ 
 small and positive at $M_\Z=0$,
 but increasing  to infinity, crossing unity on its way, as $M_\Z$ approaches 
$M+\omega_1$. This crossing point is the lowest value of $M_\Z$ that
solves \eqn{trans}). For $M_\Z$ slightly greater than $M+\omega_1$, $F(M_\Z)$ 
will rise from negative infinity and cross unity at a value between
 $M+\omega_1$  and $M+\omega_2$, which is a much higher value.
See Fig.~\ref{fig:trans1} which shows that one solution occurs at a much
lower energy than all of the others; this corresponds  $M_\Z$, with
the $\Z$  as a 
collective vibration.

The graphical solution shows that all but one of the eigenvalues
occur between the energies of the unperturbed states. A useful simplification
\cite{bb}
is to let the $\omega_j$ of \eqn{trans0})
become equal to a common value, taken as $\omega$.
Then the vertical lines of Fig.~\ref{fig:trans1} coalesce and
 all but one  the eigenvalues are equal to $\omega$. 
One value is considerably lower than the
others, the  one of the $\Z$. Then  
\bea
M_\Theta=M+\omega-{\lambda\over3} 
\sum_j\int d^3q q^2D_j^2(q).\label{trans}\eea

Since our treatment is meant to be schematic, 
we take each $D_i$ to be the same, $D_i(q)=D(q)$, and take the 
number of states to be $N (\ge18)$.
The essential feature here is that $N$ is large;
its exact value will not matter. 
 Then 
eqn.~(\ref{ceq}) can  be solved as
\bea 
c_j(q)=  \gamma D(q),
\eea
with $\gamma$ determined from the normalization of the $\Z$ as
$\gamma=1/\sqrt{N\int d^3q q^2 D(q)^2}$,
and \eqn{trans}) becomes
\bea
M+\omega-M_\Z\approx \mu={\lambda\over 3} N 
\int d^3q\;q^2D^2(q).
\label{res}\eea

The result \eqn{res}) shows the power of the coherence in
increasing the importance of the residual interaction by a 
factor of at least 18!
 This mechanism is an attractive version of the repulsive 
mechanism  used
to describe the nuclear giant dipole resonance\cite{bb}. The discovery 
of this huge collective resonance involved a puzzle.
Simple shell model considerations gave the energy of the excited
state as one unit of $\hbar\omega$, while the observed value is
$\approx 2\hbar\omega$. The coherent effect of a repulsive particle-hole
interaction was shown\cite{bb}
 to increase the energy of the giant dipole resonance. 
Here there is an attractive residual interaction between the excitations
and the core nucleon.  The replacement of $
M+\omega-M_\Z$ by $\mu$ involves noting that $\Delta E_1\approx M_\Z-M
(\approx 600)$ MeV (recall \eqn{oi})) and 
then including any effects of radial excitations in the parameter $\mu$.

We next argue that 
the wave function \eqn{wf}) leads naturally to a very weak decay
to the $KN$ channel. What are the mechanisms for 
decay? One might think of  gluon exchange, 
 but the effects of one gluon exchange are eliminated by the
color singlet nature of the nucleon and its excitation. One is left with
multi-gluon
exchanges at low energies, 
but these effects must be non-perturbative so 
it is natural to think of chiral mechanisms.
Indeed, a natural model to use when considering low-energy
excitations is the cloudy-bag model\cite{cbm}, or its 
 relativistic form\cite{Miller:2002ig}. In this model,
the pseudoscalar excitation interacts by exchanging a pion with the nucleon
core. The pion is emitted only by the $u$ or $d$ quark in the pseudoscalar
excitation, the interaction can be expressed as 
$\boldsigma\cdot \boldv$ where $\boldv$ is the relative momentum between the
pion and the light quark.
The matrix element for the transition to a  state with kaonic 
quantum numbers 
$\propto Tr\left[\sigma_2 \boldsigma\cdot
\boldv\sigma_2\right]=0,$ but the one for the transition to 
a state with K$^*$ quantum numbers is 
$\propto Tr\left[ \boldsigma\cdot
\boldv\sigma_2\right]\ne0.$   Thus transitions between the $\Z$ and nucleons 
involving  virtual $K^*$
mesons can be strong. Indeed, the absorption of 
a virtual $K^*$ by a nucleon making a $Z$ will be strongly enhanced due to
coherent collective effects.
 
The next step is to use the present model to estimate the $\Z$-nucleon
interaction. We shall see that the resulting  $\Z N$ potential will
be proportional to $-\mu$, and therefore very strong. 

\begin{figure}
\begin{center}
        \leavevmode
        \epsfxsize=0.07\textwidth
        \epsfbox{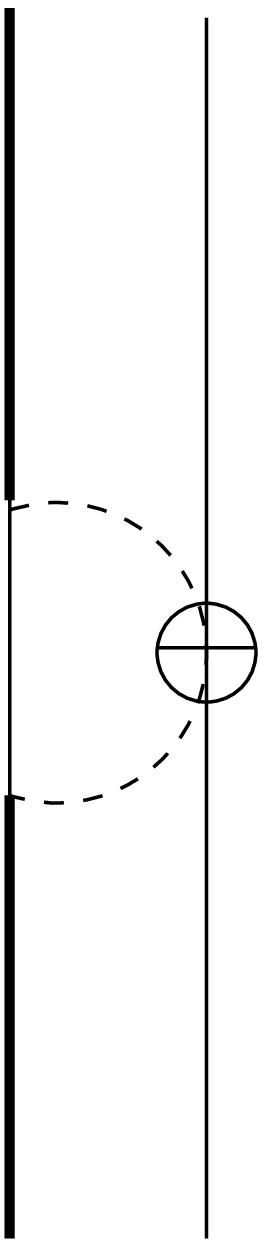}
\end{center}
\caption{\label{fig:graph } $\Theta N$ interaction. The heavy lines 
represents the $\Z$ and the light one the nucleon. Coherent 
excitations in the $\Z$ are denoted by the dashed line 
 $\widehat{V}$ is represented by the circle.}
 \end{figure}

The interaction between the $\Z$ and the nucleon is expressed in terms of
a potential. The initial $\Z N$ state is defined by the quantum numbers
$(\bfP,m_\Z;\bfp,m) $ and the final state is similarly $(\bfP',m_{\Z'};\bfp',m')
$.
The relevant matrix element is 
$\bb \Z_{{m}'_\Z}(\bfP'),N_{m'}(\bfp')\vert \widehat{V}\vert
\Z_{{m}_\Z}(\bfP),N_{m}(\bfp)\kk_c=\delta(\bfP+\bfp-\bfP'-\bfp)
\bb \Z_{{m}'_\Z}(\bfP+\bfp-\bfp'),N_{m'}(\bfp')\vert v\vert
\Z_{{m}_\Z}(\bfP),N_{m}(\bfp)\kk_c
.$ The subscript $c$ indicates that the matrix element contains a 
reproduction of the $\Z$ self-energy proportional to $V_0$ that must
be subtracted.   Evaluation of the matrix element shows that
only the  term in which the coherent cloud  of theta interacts with the
nucleon survives the evaluation\cite{exch}, 
see Fig.~\ref{fig:graph }. Then one finds
\bea
&&\bb \Z_{{m}'_\Z}(\bfP+\bfp-\bfp'),N_{m'}(\bfp')\vert v\vert
\Z_{{m}_\Z}(\bfP),N_{m}(\bfp)\kk_c=\nonumber\\
&&-\mu N\delta_{m,m'}
\delta_{m_\Z,{m'_\Z}}
\int d^3q C(q'_r)D(k'_r)C(q_r)D(k_r)\bfk'_r\cdot\bfk_r 
\bb m'_\Z\vert\boldsigma\cdot\bfq_r\;\boldsigma\cdot \bfq'_r\vert m_\Z\kk
,\label{long}\eea 
where $\bfq_r=(M\bfq-\mu
\bfP)/(M+\mu),\;\bfk_r=(M\bfq-\mu\bfp)/(M+\mu),\;
\bfq'_r=\bfq_r+\bfDelta,\;\bfk'_r=\bfk_r+\bfDelta,$
and $\bfDelta\equiv \bfP'-\bfP$.  This mechanism involves the mutual
polarization of two interacting composite quantum systems, and {\it e.g}
has  some features in common with  the two-photon exchange
interaction  responsible for  the Van der Waals force.

Let's see what \eqn{long}) tells us  about the $\Z N$ interaction.
First, simplify 
the integral over $d^3q$  by
changing variables: $\bfq\to \bfq+\mu/M \bfP\to (M+\mu)/M\bfQ,\bfQ\to
\bfQ-\bfDelta/2$. 
This allows us to re-write  \eqn{long})     as 
\bea
&&\bb \Z_{{m}'_\Z}(\bfP+\bfp-\bfp'),N_{m'}(\bfp')\vert v\vert
\Z_{{m}_\Z}(\bfP),N_{m}(\bfp)\kk_c=\delta_{m',m}
(- \mu N) \left({M+\mu\over M}\right)^3\gamma^2\times\nonumber\\
 &&
\int d^3Q
D(\vert\bfQ-\bfDelta/2\vert)D(\vert\bfQ+\bfDelta/2\vert)
D(\vert\bfQ-\bfL-\bfDelta/2\vert)D(\vert\bfQ-\bfL+\bfDelta/2\vert)
\times\nonumber\\
&&\left[\left(\bfQ-\bfL\right)^2-\bfDelta^2/4\right]
\left[\delta_{m_\Z,m'_\Z}\left(Q^2-\Delta^2/4\right)+2\bb m'_\Z\vert i\boldsigma\cdot(\bfQ\times
  \bfDelta/2)\vert {{m'}_\Z}\kk\right]
,\label{big}\eea
where $\bfL\equiv \mu/(M+\mu)(\bfp-\bfP)$. 
The quantity  $\bfp-\bfP $ is essentially the relative momentum between the
$\Z$ and the nucleon\cite{talk}. 
We'll apply this expression to compute the properties of the $\Z$ in nuclear
matter. In that case both $\bfp$ and $\bfP$ have small magnitudes on the
order of the inverse nuclear radius. Furthermore, the symmetries of the
integrand allow one to show that only terms involving powers of 
$\bfL\cdot\bfL\sim (\mu/(M+\mu))^2(R_0/R_A)^2$, and are neglected. 
Then \eqn{big}) 
 leads to an expression in which the  potential
is expressed as $\delta_{m_\Z,m'_\Z}\delta_{m',m}$ times a function
of the variable $\Delta$.  Thus, the matrix element of $\widehat{V}$
   is  equivalent to one for   
a spin-independent local potential $v(r)$, where 
 $r$ is the distance 
between the nucleon and the $\Z$. To be specific, take 
\bea
 D(K)=e^{-{\alpha K^2 R_0^2}},\eea
where $\sqrt{\alpha}R_0$ is of the order of the nucleon size $\sim$ 1 fm.
Then we obtain:
\bea
v(r)=-V_0\left(1-{1\over6}{r^2\over\alpha R_0^2}
+{r^4\over 48 \alpha^2R_0^4}\right)
\exp\left({{-r^2\over4 \alpha R_0^2}}\right),\;V_0\equiv
\mu\left({M+\mu\over M}\right)^3.\label{pot}\eea
The strength of $v(r)$ is constrained by the known mass of the $\Z$, but
also depends on the paramter $\mu$. We have argued above that
possible values of $\mu$ range between 250 and 800 MeV.
We use the lowest value to obtain
$V_0\approx 420 $ 
 MeV.
 This is amazingly strong  potential persists to relatively
long ranges on the order of a fm for any reasonable choice of $\alpha$. 
Thus there will be substantial $\Z N$ attraction, even if there is 
a  repulsive interaction at short distances 
 between the nucleon and the nucleon
constituent of the $\Z$.

The result (\ref{pot}), with its huge attraction,
 is the essential finding of this paper. It arises from the 
assumed even parity 
of the $\Z$ and the assumption that the surprisingly small  mass of 1540 MeV 
arises from a coherent interaction.

How can the influence of such an attraction be made observable?
The immediate implication is that the $\Z$ would be bound to nuclei with
a binding energy of the lowest energy state
is substantially greater than the 105 MeV threshold energy. This can
be seen by observing that the $\Z$-nucleus  mean 
field $U$ is the convolution of the interaction
 (\ref{pot}) with the density $\rho$ of nucleons within the
nucleus. Then,  with $R_\Z$ as the distance between the $\Z$ and 
 the nuclear center,
\bea U(R_\Z)=\int d^3s v(s)\rho (\bfR_\Z+\bfs)\approx \rho (R_\Z)\int d^3s v(s)
,\eea
in which the second equation arises from 
taking the nuclear radius to be much, much greater than $R_0$. Carrying out
the integral 
one finds
\bea U(R_\Z)\approx-{V_0}
  \rho(R_\Z){10}(\alpha\pi)^{3/2}.\eea
To estimate the central maximum value take $\rho(0)$ to be the density
of infinite nuclear matter, $(1/6){\rm fm}^{-3}$ and $\alpha=1/4$ so that the
exponential term 
 of the interaction $v(r)$ has a range of  $R_0=1$ fm. This gives the central
nuclear potential a value of about 490 MeV.
 This is 
 a huge attractive potential. For such a deep potential, the binding
energy is close to the central value of the potential.
This is much larger than the 105 MeV threshold energy,
so the $\Z$ bound to the nucleus will be stable against decay by
strong interaction effects. 
A more sophisticated calculation would include the repulsive influence of 
nucleon-nucleon correlations and reduce the strength of the attractive 
interaction. However, with the mean field calculation  giving  a binding energy
 of 490 MeV, we can be sure that a huge attraction survives.
Thus we predict that 
a new state of nuclear
matter, $\Z$-matter of positive strangeness and excitation
 energy of the order of hundreds
of MeV exists.

How can $\Z$-matter be detected? It has long been known that
hypernuclei can be made in reactions in which a nucleon is replaced
by a hyperon of roughly the same momentum\cite{Bruckner:1974qy}.
 With this in mind, it's a  straightforward exercise in kinematics 
 to see that 
$\Z$ nuclei can be made using photon or kaon beams of energies from about 1 to
5 GeV or more.
Denote $(A-1)_Z$ as a state
of baryon number $A-1$ containing one $\Z$. Then  reactions
$\gamma +A\to \Sigma+ (A-1)_\Z,K+A\to \pi+(A-1)_\Z$, are prime candidates
for reactions that would 
copiously produce $\Z$-nuclei. 
This  because the required transfer to the nucleus is
very small.  For example, for a $^{40}$Ca target, $\Z$ binding energy
of 200 MeV, and a photon beam of 3 GeV, $(p_\gamma-p_\Sigma)^2
=.004\; {\rm GeV}^2$ for forward production.
 Similarly a kaon beam of 3 GeV would have $(p_K-p_\pi)^2
=.026 \;{\rm GeV}^2$. 

There are  other 
implications of the present model. Since the low mass of the $\Z$
is caused by a non-confining residual interaction involving a
$u\bar s$ or $d\bar s$ pair, it is reasonable to expect
that   
$u\bar c$,  $u\bar c$  and $d\bar b$,$d\bar c$ will interact with
nucleons in a similar manner. Thus there should be
a charmed pentaquark at about 1540 MeV plus the $c$-$s$ quark mass difference
of about 1000 MeV. Such state would be stable 
against strong decay into a D meson and a nucleon.
Similarly, the bottom pentaquark system would be stable against decay into 
a a B meson and a nucleon.
Thus, in agreement with previous 
authors\cite{Jaffe:2003sg}, we expect that 
the charmed and bottom versions of the pentaquark will be stable against
particle decay. 

The present schematic model is too naive for detailed applications to
spectroscopy, which generally is a difficult subject to pursue\cite{quibble}.
However, discussing the energy of the doubly strange
version, a possible $\Sigma(3/2)$,
  of the pentaquark  observed in Ref.~\cite{Alt:2003vb} is worthwhile.
 If the
doubly strange system  and the $\Z$ really are members of 
the same multiplet
then we may estimate $M_{\Xi(3/2)}$ as 
$M_{\Xi(3/2)}=M_\Z+(M_s-m_d) +(E_{3/2}-E_{1/2})$ in which the
The term
in parenthesis is an estimate of the influence of the  difference 
in angular momentum. Taking this from the mass difference between
doubly strange $\Xi(1530)$ J=3/2 and $\Xi(1320)$ J=1/2 states
gives $M_{\Xi(3/2)}
\approx 1920$ MeV, which 
 is in fair 
 agreement with the experimental value of 1862 MeV.

We have presented a schematic model of quark-pair interactions with
nucleons that reproduces the essential features of an even parity
strange pentaquark. The  attractive
schematic interaction 
gets a huge strength from collective
coherent effects and therefore reproduces the 
low mass (1540 MeV)
small width (to the $KN$  channel) but does not lead to a suppression 
of the production of a virtual kaon, $K^*$. In this model,
 the $\Z$ can be regarded as a
collective vibration of the nucleon.  Determining the fundamental
origin  of the schematic interaction would be a task for further work.

A natural consequence of the strong
attraction is that $\Z$ nuclei, stable against strong decay, 
may exist. Such states  can be made in photon and kaon beam experiments.
Suppose the $\Z$ has odd parity. Then one need not account for the
excitation energy of 600 MeV, using an attractive residual 
interaction, and one would not predict that $\Z$-nuclear matter would be stable
against decay by the strong interaction. Therefore, the discovery
of such stable $\Z$-nuclear matter could be a definitive proof that the
$\Z$ parity is even.

\begin{acknowledgments}

This work  owes its genesis to  a conversation with M. Polyakov.
I thank D. Ashery, P. Bedaque, V. Koch,  and 
E. Piazetsky for useful discussions. 
I am  grateful to the Nuclear Science Theory Group of 
Lawrence Berkeley Lab,  ECT* (Trento), the 
  National Institute for Nuclear Theory at the 
University of Washington 
 and the CSSSM (Adelaide) 
for providing hospitality during the course of carrying out this
work. The Guggenheim foundation, on the other hand, was of no help at all.
This work is partially  supported by 
the USDOE grant DE-FG-02-97ER41014.
\end{acknowledgments}

\bibliographystyle{unsrt}

\begin{thebibliography}{99}
\bibitem{exch} There are exchange terms which occur if
$\bfP=\bfP'$, but this corresponds to a set of zero measure and
does not contribute to scattering.
\bibitem{talk} The expression for the interaction matrix element
\eqn{big}) is invariant under the transformation $\bfq\to\bfq+\mu
\boldv,\;\bfp\to\bfp+M\boldv,\;\bfP\to\bfP+M\boldv$.  For a weakly bound
system the difference between the mass of the nucleon consituent and
$M_\Z$ would be small, so that $\bfp-\bfP$ would be the correct
relative momentum. Here we have strong binding 
 caused by the effects of the residual interaction.
 Keeping such effects  in the transformation between frames 
is a relativistic effect and is  
ignored here.
\end{thebibliography}

\end{document}